\documentclass[aip,cha, preprint,showpacs,showkeys]{revtex4-1}
\usepackage{amssymb, dcolumn,  graphicx, latexsym, amsfonts, bm}

\begin{document}

\title{The nonmodal kinetic theory for the electrostatic instabilities of a plasma with a 
sheared Hall current.}
\author{V. V. Mikhailenko}\email[E-mail: ] { vladimir@pusan.ac.kr}
\affiliation{Plasma Research Center, Pusan National University, Busan 46241, South Korea.}
\author{V. S. Mikhailenko}\email[E-mail:] {vsmikhailenko@pusan.ac.kr}
\affiliation{Plasma Research Center, Pusan National University, Busan 46241, South Korea.}
\author{H. J. Lee}\email[E-mail: ] {haejune@pusan.ac.kr}%
\affiliation{Department of Electrical Engineering, Pusan National University, Busan 46241, 
South Korea.}

\begin{abstract}
The kinetic theory for the instabilities driven by the Hall current with a 
sheared current velocity, which has the method of the shearing modes or the so-called non-
modal approach as its foundation, is developed.
The developed theory predicts that in the Hall plasma with the
inhomogeneous electric field, the separate spatial Fourier mode of the perturbations is 
determined in the frame convected with one of the plasma 
components. Because of the different shearing of the ion and electron flows in the Hall 
plasma, this mode is perceived by the second component as the 
Doppler-shifted continuously sheared mode with time-dependent wave numbers. Due to this 
effect, the interaction of the  plasma components forms the nonmodal time-dependent process, 
which should be investigated as the initial value problem. The developed approach is applied 
to the solutions of the linear initial value problems for the hydrodynamic modified two-
stream instability and the kinetic ion-sound instability of the plasma with a sheared Hall 
current with a uniform velocity shear. These 
solutions  reveal that the uniform part of the current velocity  is responsible for 
the modal evolution of the instability, whereas the current velocity shear is the source of 
the development of the nonmodal instability with exponent growing with time as $\sim\left(t-
t_{0}\right)^{3}$.
\end{abstract}


\maketitle

\section{Introduction}\label{sec1}

The crossed electric, $\mathbf{E}$, and magnetic, $\mathbf{B}$, fields configuration is 
frequently observed in fusion\cite{Burrell} and 
space \cite{Mozer} plasmas. This configuration is common to a large variety of the so-called 
$E\times B$ plasma devices\cite{Boeuf}, 
which includes the Hall-thrusters\cite{Morozov,Goebel}, cylindrical and planar magnetrons
\cite{Anders}, and numerous other applications. 
Depending on the plasma and fields parameters, plasma response on these fields is very 
different. 

The flow of the collisionless plasma with strongly magnetized electrons and ions forms in the 
spatially homogeneous crossed $\mathbf{E}$, 
$\mathbf{B}$ fields. The velocity of this flow, $V_{0}=cE/B$, is the same for all particles 
such that this flow produces no current in a 
charge-neutral plasma and does not have any effects on the plasma stability. However, the 
sheared 
$\mathbf{E}\times\mathbf{B}$ poloidal rotation of the tokamak 
edge with spatially inhomogeneous electric field suppresses the instabilities in the drift - 
frequency range\cite{Burrell}, which are 
responsible for the anomalous transport of plasma. This process  is of critical importance 
for the formation and control of the high confinement mode of operation, or H-mode.

Contrary to the fusion plasmas, the Larmor radius of ions in the $\mathbf{E}\times\mathbf{B}$ 
devices is not small with respect to the dimensions of the system and the ions are considered 
as unmagnetized, while electrons are strongly magnetized. The relative motion of the 
unmagnetized ions and the strongly magnetized electrons drifting 
with velocity $V_{0}=cE/B$ in such a plasma (generally referred to as the Hall plasma) forms 
the Hall current. This specific current, which is absent in plasmas with  all magnetized
species, is the source of numerous current-driven instabilities which have been observed 
experimentally\cite{Tilinin,Choueiri,Lazurenko,Tsikata,Litvak1} and 
in simulations\cite{Chable,Fernandez,Matyash,Boeuf 2}, and were investigated analytically
\cite{Simon,Hoh,Sakawa,
Lashmore-Davies,Gary1,Gary2,Litvak2,Cavalier,Smolyakov,Romadanov,Koshkarov,Frias1,Frias2}. 
The discovered instabilities are very dependent on the specific conditions and regions of a 
particular device and develops in a large range of frequencies and wavelengths which includes 
large scale low frequency 'rotating spokes' \cite{Parker,Boeuf 1,Rodríguez,Matyash}, the 
modified two-stream (MTS) instability and ion sound (IS) and lower hybrid instabilities with 
frequencies between the ion cyclotron and electron cyclotron frequencies, and the
submillimeter electron cyclotron drift instabilities\cite{Cavalier,Janhunen} in the MHz 
frequency range. Important set of the Hall plasma instabilities is the gradient-drift plasma 
instabilities 
\cite{Simon,Hoh,Sakawa,Smolyakov,Romadanov,Koshkarov,Litvak2,Frias1,Frias2,Romadanov} which 
develop in spatially inhomogeneous Hall plasma due to the 
combined effect of the Hall current and of the gradient-drift modes formed by the plasma 
density and temperature inhomogeneity. A typical example of these instabilities are 
the Simon-Hoh instabilities\cite{Simon,Hoh,Sakawa,Koshkarov} which have frequency much above 
the ion cyclotron frequency, but below the electron cyclotron frequency. It is generally 
believed that turbulence powered by the instabilities  
\cite{Lafleur1,Lafleur2,Lafleur3} is responsible for anomalous transport of the Hall plasma 
and is considered as a source of the experimentally detected anomalous electron mobility
\cite{Meezan} in Hall plasma thrusters. 

The stability theory of the Hall plasmas historically bases on the normal mode analysis. It 
successfully identifies the waves and instabilities in the Hall plasmas by employing the 
local approximation and modal approach for the homogeneous or weekly inhomogeneous plasma, 
which assumes that the plasma perturbations have a structure of a plane wave $\sim \exp
\left(i\mathbf{kr}-i\omega t\right)$. In the case of the spatially inhomogeneous plasma, the 
nonlocal analysis of plasma stability is performed\cite{Romadanov} assuming that the 
perturbations have a form $\sim \phi\left(x\right)\exp\left(ik_{y}y+ik_{z}z-i\omega t\right)$ 
in the nonuniform along coordinate $x$ plasma and solving the eigenfunction-eigenvalue 
problem for the mode structure $\phi\left(x\right)$ and frequency $\omega$.  

Generally, the electric and magnetic fields in the $E\times B$ devices are spatially 
inhomogeneous, and the corresponding Hall current is spatially inhomogeneous and sheared. It 
was found in Ref.\cite{Mikhailenko1}, where the hydrodynamic theory of the modified 
Simon-Hoh (MSH) instability of a plasma with a sheared Hall current was developed, that the 
local approximation, which admits the 
application of the modal plane wave approach to the stability analysis of the Hall plasma, 
should be revised when it applies to the sheared Hall current. The solution of the initial 
value problem in Ref.\cite{Mikhailenko1}, instead of application of the spectral transform in 
time, discovered  nonmodal exponential growth of the perturbations with time as $\sim\exp a
\left(t-t_{0}\right)^{3}$ for this instability. This growth is missed in the normal mode 
analysis. It was found that this nonmodal growth 
dominates the normal mode growth when the current velocity shearing rate is 
above the growth rate of the  MSH instability  and  includes also the nonmodal growth of the 
perturbations which are subcritical for the MSH instability of the 
plasma with uniform Hall current. This result confirms the general conclusion derived earlier 
in Refs.\cite{Trefethen, Schmid, Farrell, Mikhailenko-2000, Schmid2, Camporeale, Camporeale1, 
Friedman, Squire,  Mikhailenko2, Mikhailenko3}, that the nonmodal effects of the 
{\textit{sheared flow}} as well as the 
derived in Ref.\cite{Mikhailenko1} nonmodal effects of the {\textit{sheared current}}, are 
missed completely in the usual normal modes analysis and 
the investigation of the stability of  sheared flows and sheared currents needs more 
elaborated analysis grounded on the methodology of the sheared modes and solution of the 
corresponding initial value problems. 

Kinetic effects, such as finite-Larmor-radius effects, the Landau and cyclotron damping, and 
existence of numerous kinetic instabilities, which are naturally not involved in the
fluid description of plasma shear flows, require the development
of a kinetic theory of stability of the Hall plasma with a sheared Hall current. 
In this paper, we present an analytical nonmodal approach grounded on the methodology of 
shearing modes to the kinetic theory of instabilities driven by the sheared current. The 
governing equations of this theory for the electrostatic instabilities are derived in details 
in Sec. \ref{sec2}. The application of the developed approach to the theory of the MTS and of 
the kinetic IS instabilities of a plasma with a sheared Hall current are given in Secs. 
\ref{sec3} and \ref{sec4}. Conclusions are given in Sec. \ref{sec5}.

\section{The shearing modes approach to the theory of the instabilities driven by the 
sheared current}\label{sec2}

Our theory bases on the system of Vlasov equations for electrons and ions and the Poisson 
equation for the perturbed electrostatic potential. In this paper, we consider a plasma in 
the linearly changing electric field $\mathbf{E}_{0}\left( \mathbf{r}\right)= \mathbf{E}_{0}
\left( \mathbf{r}=0\right)+ E'_{0}x\mathbf{e}_{x}$ with  
$E'_{0}=\partial E_{0}/\partial x=const$, directed across the 
uniform magnetic field $\mathbf{B}=B
\mathbf{e}_{z}$ pointed along the coordinate $z$. The strength 
of the magnetic field is such that the ion Larmor radius $\rho_{i}$ is much larger than the 
characteristic plasma length scale $L$, 
whereas the electron Larmor radius $\rho_{e}$ is much less than $L$. We will consider the 
electrostatic perturbations with a frequency 
much above the ion cyclotron frequency $\omega_{ci}$. The evolution of the magnetized 
electrons is governed by the Vlasov equation for the electron distribution function $F_{e}
\left(\mathbf{v}, \mathbf{r}, t\right)$,
\begin{eqnarray}
&\displaystyle \frac{\partial F_{e}}{\partial
t}+\mathbf{v}\frac{\partial F_{e}}
{\partial\mathbf{r}}+\frac{e}{m_{e}}\left(\left(E_{0}+E'_{0}x\right)\mathbf{e}
_{x}\right.
\nonumber
\\ &\displaystyle \left.+\frac{1}{c}\left[\mathbf{v}\times\mathbf{B_{0}}\right]-\nabla 
\varphi\left(\mathbf{r}, t\right)\right)
\frac{\partial F_{e}}{\partial\mathbf{v}}
= 0.\label{1}
\end{eqnarray}
The evolution of the unmagnetized ions is governed by the Vlasov equation for 
$F_{i}\left(\mathbf{v}, \mathbf{r}, t\right)$,
\begin{eqnarray}
&\displaystyle \frac{\partial F_{i}}{\partial
t}+\mathbf{v}\frac{\partial F_{i}}
{\partial\mathbf{r}}+\frac{e}{m_{i}}\left(\left(E_{0}+E'_{0}x\right)\mathbf{e}
_{x}-\nabla \varphi\left(\mathbf{r}, t\right)\right)
\frac{\partial F_{i}}{\partial\mathbf{v}}
= 0.\label{2}
\end{eqnarray}
The potential $\varphi\left(\mathbf{r}, t\right)$ in Eqs. (\ref{1}) and (\ref{2}) is 
determined by the Poisson equation,
\begin{eqnarray}
&\displaystyle 
\vartriangle \varphi\left(\mathbf{r},t\right)=
-4\pi\sum_{\alpha=i,e} e_{\alpha}n_{\alpha}\left(\mathbf{r},t\right)
\nonumber
\\ &\displaystyle
=-4\pi\sum_{\alpha=i,e} e_{\alpha}\int f_{\alpha}\left(\mathbf{v},
\mathbf{r}, t \right)d\mathbf{v}, 
\label{3}
\end{eqnarray}
where $f_{\alpha}\left(\mathbf{v}, \mathbf{r}, t \right)=F_{\alpha}\left(\mathbf{v}, 
\mathbf{r}, t \right)-F_{0\alpha}\left(\mathbf{v}\right)$ is the 
perturbation of the electron $\left(\alpha=e\right)$  and ion $\left(\alpha=i\right)$ 
distribution functions. The simplest solutions to Eqs. (\ref{1}), (\ref{2}) may be obtained 
applying the so called local approximation. 
Usually, when the local approximation supposed to apply to the Vlasov equation (see, for 
example, Ref.\cite{Artun}) for the magnetized electrons, the transformation of Eq.
(\ref{1})  to the frame of references that moves with  velocity $\mathbf{V}_{e0}
\left(\mathbf{r}\right)=c\mathbf{E}_{0}\left(x\right)\times\mathbf{B}/B^{2}$ in the electron 
velocity space, but unchanged in the configuration space, is employed. Also, the 
transformation to the frame of references that 
moves with  velocity $\mathbf{V}_{i0}\left(\mathbf{r}\right)\approx\left(e/m_{i}\right)
\mathbf{E}_{0}t$ in the ion velocity space, but unchanged in the configuration space, we 
employ to  Eq.(\ref{2}) for the unmagnetized ions. With new velocities $\mathbf{v}_{e}= 
\mathbf{v}-\mathbf{V}_{e0}\left(\mathbf{r}\right)$ and $\mathbf{v}_{i}= \mathbf{v}- 
\mathbf{V}_{i0}\left(\mathbf{r}\right)$ Eqs. (\ref{1}) and (\ref{2}) become
\begin{eqnarray}
&\displaystyle \frac{\partial F_{e}}{\partial
t}+V_{e0}\left(x\right)\frac{\partial F_{e}}
{\partial y}+\mathbf{v}_{e}\frac{\partial F_{e}}
{\partial \mathbf{r}}
\nonumber 
\\
&\displaystyle
+\frac{e}{m_{e}}\left(\frac{1}{c}\left[\mathbf{v}_{e}\times\mathbf{B}\right]-\nabla \varphi
\left(\mathbf{r},t\right)\right)\frac{\partial F_{e}}{\partial\mathbf{v}_{e}}=0,\label{4}
\end{eqnarray}
\begin{eqnarray}
&\displaystyle \frac{\partial F_{i}}{\partial t}+V_{i0}\left(x\right)\frac{\partial F_{i}}
{\partial y}+\mathbf{v}_{i}\frac{\partial F_{i}}{\partial \mathbf{r}}
-\frac{e}{m_{i}}\nabla \varphi\left(\mathbf{r},t\right)\frac{\partial F_{i}}{\partial
\mathbf{v}_{i}}=0,\label{5}
\end{eqnarray}
where it was assumed in Eq. (\ref{4}) that the electron flow velocity shear $V'_{e}=dV_{e0}
\left(x\right)/dx 
=-cE'_{0}/B= const$ is much less than the electron cyclotron frequency $\omega_{ce}$.
The local approximation grounds on the assumption that all modes being considered have  
wavelengths significantly shorter than the spatial scale length $L_{V_{e,i}}$ of 
$V_{e0,i0}\left(x\right)$ velocities inhomogeneities, i.e. 
\begin{eqnarray}
&\displaystyle 
L_{V_{e,i}}k_{x}\gg 1.
\label{6}
\end{eqnarray}
With local approximation, the solutions  of both Vlasov equation for the perturbations $f_{e}
$ and $f_{i}$ of the 
equilibrium electron distribution functions $F_{e0}$ and $F_{i0}$ are derived in the modal 
form of a plane 
wave $\sim \exp\left(i\mathbf{kr}-i\omega t\right)$, considering velocities $V_{e0, i0}
\left(x\right)$ as spatially 
homogeneous. The employment of the obtained solutions $f_{e}\left(\mathbf{v}_{e}, 
\mathbf{k}, \omega-k_{y}V_{e0}\right)$ and $f_{i}\left(\mathbf{v}_{i}, \mathbf{k}, \omega-
k_{y}V_{i0}\right)$ in the Fourier transformed Poisson equation gives the well known local 
dispersion equation 
\begin{eqnarray}
&\displaystyle 
1+\varepsilon_{e}\left(\mathbf{k}, \omega-k_{y}
V_{e0}\left(x\right)\right)
+ \varepsilon_{i}\left(\mathbf{k}, 
\omega-k_{y}V_{i0}\left(x\right)\right)=0
\label{7}
\end{eqnarray}
as for the instabilities driven by the spatially uniform current with current velocity 
$U=V_{e0}-V_{i0}$. We found\cite{Mikhailenko1}, however, that the condition (\ref{6}) of the 
local approximation is not sufficient for the application of the modal approach to the 
stability analysis of the plasma with a sheared Hall current. In this paper, we derive the 
solutions to the Vlasov-Poisson system (\ref{4}), (\ref{5}) for $f_{e}$, $f_{i}$ and 
(\ref{3}) as 
of the initial value problem without application of the spectral transforms over time 
variable. On this way, we will find the above mentioned solutions of the modal type, and 
derive additional conditions which are necessary for the validity of the modal solutions. 
Also, we find the nonmodal solutions, which are missed in the conventional normal mode 
analysis.

The first step of our approach to the solution of the system (\ref{4}), (\ref{5}), and 
(\ref{3}) 
is the transformation of the spatial coordinates $\mathbf{r}$ in Eqs. (\ref{4}) for $F_{e}$, 
determined in the laboratory frame, to the coordinates $\mathbf{r}_{e}$ determined in the 
frame moving with velocity $\mathbf{V}_{e0}\left(x\right)=c
\left(E_{0}+E'_{0}x\right)\mathbf{e}_{y}/B=
\left(V_{0}+V'_{0}x\right)\mathbf{e}_{y}$ of a sheared equilibrium electron flow with uniform 
velocity shear $V'_{0}$. With coordinates $x_{e}, y_{e}, z_{e}$ 
and velocities $v_{ex}, v_{ey}, v_{ez}$ determined in the convected electron frame by the 
relations
\begin{eqnarray}
&\displaystyle
x=x_{e}, \quad y=y_{e}+V_{e0}t+V'_{e}x_{e}t, \quad z=z_{e},
\nonumber
\\ &\displaystyle 
v_{x}=v_{ex}, \quad v_{y}=v_{ey}+ V_{e0}+V'_{e}x_{e},\quad v_{z}=v_{ez},
\label{8}
\end{eqnarray}
where it was assumed that the electric field $\mathbf{E}_{0}$ emerges at time $t=0$, Eq. 
(\ref{4}) 
becomes
\begin{eqnarray}
&\displaystyle \frac{\partial F_{e}}{\partial t}+v_{ex}\frac{\partial F_{e}}{\partial x_{e}} 
+\left(v_{ey}-v_{ex}V'_{e}t \right) \frac{\partial
F_{e}}{\partial y_{e}} +\omega_{ce} v_{ey}
\frac{\partial F_{e}}{\partial v_{e
x}}-\omega_{ce}v_{ex}\frac{\partial F_{e0}}{\partial v_{ey}} \nonumber
\\  &\displaystyle
-\frac{e}{m_{e}}\left(\frac{\partial \varphi}{\partial
x_{e}} -V'_{e}t\frac{\partial \varphi}{\partial y_{e}} \right)
\frac{\partial F_{e}}{\partial v_{ex}}+v_{ez}\frac{\partial F_{e}}{\partial z_{e}}
-\frac{e}{m_{e}} \frac{\partial \varphi}{\partial y_{e}}
\frac{\partial F_{e}}{\partial v_{ey}}
-\frac{e}{m_{e}} \frac{\partial \varphi}{\partial
z_{e}} \frac{\partial F_{e}}{\partial v_{ez}}=0.\label{9}
\end{eqnarray}
The explicit spatial inhomogeneity introduced by the electric field $\mathbf{E}_{0}
\left(x\right)$ is absent in the Vlasov equation (\ref{9})\cite{Mikhailenko2}. With electron 
guiding center coordinates $X_{e}$, $Y_{e}$, determined in the electron convective frame, 
\begin{eqnarray}
&\displaystyle
x_{e}=X_{e}-\frac{v_{e\bot}}{\omega_{ce}}\sin \left(\phi_{1}-\omega_{ce}t\right),
\nonumber
\\ &\displaystyle
y_{e}=Y_{e}+\frac{v_{e\bot}}{\omega_{ce}}\cos \left(\phi_{1}-\omega_{ce}t\right)+V'_{e}t
\left(X_{e}-x_{e}\right),
\nonumber
\\ &\displaystyle
z_{e}=z_{e1}+v_{ez}t,
\label{10}
\end{eqnarray}
where $\phi=\phi_{1}-\omega_{ce}t, \quad v_{ex}=v_{e\bot}\cos \phi, \quad v_{ey}=v_{e\bot}
\sin \phi$, Eq. (9) has the most simple form
\begin{eqnarray}
&\displaystyle \frac{\partial F_{e}}{\partial
t}+\frac{e}{m_{e}\omega_{ce}}
\left(\frac{\partial\varphi}{\partial X_{e}} \frac{\partial
F_{e}} {\partial Y_{e}}-\frac{\partial\varphi}{\partial
Y_{e}} \frac{\partial F_{e}} {\partial X_{e}}\right)
\nonumber \\
&\displaystyle +\frac{e}{m_{e}}\frac{\omega_{ce}}{v_{e\bot}}
\left(\frac{\partial\varphi}{\partial \phi_{1}} \frac{\partial
F_{e}} {\partial v_{e\perp}}-\frac{\partial\varphi}{\partial
v_{e\perp}}\frac{\partial
F_{e}} {\partial \phi_{1}}\right)
-\frac{e}{m_{e}}\frac{\partial\varphi}{\partial
z_{e}} \frac{\partial F_{e}}{\partial v_{ez}}=0. \label{11}
\end{eqnarray}
The potential $\varphi$ in Eqs. (\ref{9}), (\ref{11}) is determined in the same electron 
convective-sheared coordinates (\ref{8}) or (\ref{10}), and may be presented by the Fourier 
transform as 
\begin{eqnarray}
&\displaystyle \varphi \left(x_{e}, y_{e}, z_{e},
t\right) = \int\varphi_{e}\left(k_{ex}, k_{ey}, k_{ez} ,t
\right)e^{ik_{ex}x_{e}
+ik_{ey}y_{e}+ik_{ez}z_{e}} dk_{ex}dk_{ey}dk_{ez}\nonumber  \\
&\displaystyle = \int\varphi \left(k_{ex}, k_{ey}, k_{ez}, t \right)
\exp \left[ ik_{ex}X_{e}+ik_{ey}Y_{e}+ik_{ez}z_{e}\right.  \nonumber  \\
&\displaystyle \left. -i\frac{k_{e\bot}\left(t\right)
v_{e\bot}}{\omega_{ce}}\sin\left( \phi-\omega_{ce}t -\theta
\left(t\right)\right)\right]dk_{ex}dk_{ey}dk_{ez},\label{12}
\end{eqnarray}
where
\begin{eqnarray}
&\displaystyle
k^{2}_{e\perp}\left(t\right)=\left(k_{ex}-V'_{e}tk_{ey}\right)^{2}+k_{ey}^{2},\label{13}
\end{eqnarray}
and $\tan \theta =k_{ey}/\left(k_{ex}-V'_{e}tk_{ey}\right)$.  The subscripts in $\varphi_{e}
\left(\mathbf{k}_{e},t\right)$  denotes that the potential and its Fourier transform are 
determined in the electron convected frame. 

It was obtained in Ref.\cite{Mikhailenko2}, that the equilibrium distribution function 
$F_{e0}$, which in laboratory frame contains the spatial inhomogeneity resulted from electric
field $\textbf{E}_{0}\left(\textbf{r}\right)$, does not contain such
inhomogeneity in convective coordinates (see Appendix 1 in Ref.\cite{Mikhailenko2}). In what
follows, we consider the equilibrium distribution function $F_{e0}$
as a Maxwellian,
\begin{eqnarray}
& \displaystyle F_{e0}=\frac{n_{e0}}{\left(2\pi v^{2}_{Te}\right)^{3/2}}
\exp\left(-\frac{v^{2}_{e\bot}+v^{2}_{ez}}{v^{2}_{Te}} \right).
\label{14}
\end{eqnarray}
Therefore, the Vlasov equations  (\ref{9}) or (\ref{11}) for $f_{e}\left(\mathbf{v}_{e}, 
\mathbf{r}_{e}, t \right)$ does not contain the spatial inhomogeneity in the explicit form. 
These equations may be Fourier transformed over coordinates $x_{e}, y_{e}, z_{e}$ with 
their conjugate wave numbers
$k_{ex}$, $k_{ey}$ and $k_{ez}$ without any limitations imposed by the local approximation. 
Then, the equation for the separate spatial Fourier mode $f_{e}\left(\mathbf{v}_{e}, 
\textbf{k}_{e},t\right)$ of the perturbation of the
distribution function is derived as a function of the separate Fourier mode  $\varphi
\left(\textbf{k}_{e},t\right)$ of the electrostatic potential. The solution to  Eq. 
(\ref{11}) for $f_{e}\left(\mathbf{v}_{e}, \textbf{k}_{e},t\right)$ is 
calculated easily for any magnitudes of the velocity shear rate $V'_{e}$ and is equal to
\begin{eqnarray}
&\displaystyle f_{e}\left(t,\mathbf{k}_{e},
v_{e\bot},\phi,v_{ez}\right)=\frac{ie}{m_{e}}\sum\limits_{n=-\infty}^{\infty}
\sum\limits_{n_{1}=-\infty}^{\infty} \int\limits_{t_{0}}^{t}dt_{1}
\varphi\left(t_{1},\mathbf{k}_{e}\right) \nonumber  \\
&\displaystyle\times
\exp\Big(-ik_{ez}v_{ez}\left(t-t_{1}\right)+in\left(
\phi_{1}-\omega_{ce}t-\theta\left(t\right)\right)-in_{1}\left(
\phi_{1}-\omega_{ce}t_{1}-\theta\left(t\right)\right) \Big)\nonumber
\\ &\displaystyle\times
J_{n}\left(\frac{k_{e\bot}\left(t\right)v_{e\bot} }
{\omega_{ce}}\right) J_{n_{1}}
\left(\frac{k_{e\bot}\left(t_{1}\right)v_{e\bot} } {\omega_{ce}}\right)
\left[\frac{k_{ey}}{\omega_{ce}} \frac{\partial
F_{e0}}{\partial X_{e}}+ \frac{\omega_{ce}n_{1}}{v_{e\bot}}
\frac{\partial F_{e0}}{\partial v_{e\bot}}+ k_{ez}\frac{\partial
F_{e0}}{\partial v_{ez}} \right], \label{15}
\end{eqnarray}
where $t_{0}\geq 0$ is the initial time. In the electron convective coordinates, the Fourier 
transform $n_{e}\left(\mathbf{k}_{e},t\right)$ of the perturbed electron 
density is the separate spatial Fourier mode
\begin{eqnarray}
&\displaystyle
n^{(e)}_{e}\left(\mathbf{k}_{e},t\right) =\int d\mathbf{r}_{e}n_{e}\left(\mathbf{r}_{e},t
\right)e^{-i\mathbf{k}_{e}\mathbf{r}_{e}}
\nonumber
\\ &\displaystyle 
= \int f_{e}\left(\mathbf{v}_{e}, \mathbf{k}_{e},t\right)d\mathbf{v}_{e},
\label{16}
\end{eqnarray}
where the subscript in $n^{(e)}_{e}$  denotes the  electron perturbed density, and 
superscript denotes that its Fourier 
transform is calculated in the electron convected  frame. For the equilibrium Maxwellian 
electron distribution (\ref{14}), $n^{(e)}_{e}\left(\mathbf{k}_{e},t\right)$ is equal to
\begin{widetext}
\begin{eqnarray}
&\displaystyle
n^{(e)}_{e}\left(\mathbf{k}_{e},t\right) =-\frac{2\pi e n_{e0}}{T_{e}}\sum \limits^{\infty}
_{n=-\infty}\int\limits^{t}_{t_{0}}dt_{1}\varphi_{e}
\left(\mathbf{k}_{e},t\right)I_{n}\left(k_{e\bot}\left(t\right)k_{e\bot}\left(t_{1}\right)
\rho_{e}^{2}\right)
\nonumber
\\ &\displaystyle 
\times\exp\left[-\frac{\rho^{2}_{e}}{2}\left(k^{2}_{e\bot}\left(t\right)+k^{2}_{e\bot}
\left(t_{1}\right)\right)
-\frac{1}{2}k^{2}_{ez}v^{2}_{Te}\left(t-t_{1}\right)^{2}-in\omega_{ce}\left(t-t_{1}\right)-in
\left(\theta_{e}\left(t\right)-\theta_{e}
\left(t_{1}\right)\right)\right]
\nonumber
\\ &\displaystyle \times\left(in\omega_{ce}+ k^{2}_{ez}v^{2}_{Te}\left(t-t_{1}\right)\right).
\label{17}
\end{eqnarray}
\end{widetext}
In Eq. (\ref{17}), $I_{n}$ is the modified Bessel function of the first kind and order $n$, 
$k^{2}_{e\bot}\left(t\right)= \left(k_{ex}-
k_{ey}V'_{0}t\right)^{2}+k_{ey}^{2}+k_{ez}^{2}$, $\sin \theta\left(t\right)=k_{ey}/k_{e\bot}
\left(t\right)$, $\rho_{e}=v_{Te}/\omega_{ce}$ is the thermal 
electron Larmor radius, $v_{Te}$ is the electron thermal velocity.

The model of Hall plasmas with unmagnetized ions and magnetized electrons is applicable to
the processes whose temporal evolution is limited by the
time much less then the period $\omega_{ci}^{-1}$ of the ion cyclotron Larmor rotation.
At the time interval $t-t_{0}\ll \omega_{ci}^{-1}$, the accelerated velocity $\mathbf{V}_{i0}
\left(\mathbf{r}\right)\approx\left(e/m_{i}\right)\mathbf{E}_{0}t$ of the 
unmagnetized ion component in the electric field $\mathbf{E}_{0}$ 
is much less\cite{Mikhailenko1} than the electron flow velocity $\mathbf{V}_{0}\left(x\right)
$. Therefore, for this time 
interval, we can neglect influence by the electric 
field $\mathbf{E}_{0}$, as well as by the magnetic field, $\mathbf{B}$, in the Vlasov 
equation for ions and identify the ion frame with 
a laboratory frame. The  perturbed ion density $n_{i}\left(\mathbf{r}_{i},t \right)=\int 
f_{i}\left(\mathbf{v}_{i}, \mathbf{r}_{i},t 
\right)d\mathbf{v}_{i}$ is calculated in the ion (laboratory) frame employing the ion Vlasov 
equation for the perturbation $f_{i}
\left(\mathbf{v}_{i}, \mathbf{r}_{i},t \right)$ of the ion distribution function $F_{i0}
\left(\mathbf{v}_{i}\right)$, 
\begin{eqnarray}
&\displaystyle \frac{\partial f_{i}}{\partial
t}+\mathbf{v}_{i}\frac{\partial f_{i}}
{\partial\mathbf{r}_{i}}=\frac{e}{m_{i}}\nabla \varphi\left(\mathbf{r}_{i}, t\right)
\frac{\partial
F_{i0}\left(\mathbf{v}_{i}\right)}{\partial\mathbf{v}_{i}}.\label{18}
\end{eqnarray}
For the ion equilibrium Maxwell distribution $F_{i0}\left(\mathbf{v}_{i}\right)$, 
\begin{eqnarray*}
&\displaystyle F_{i0}\left(\mathbf{v}_{i}\right)=\frac{n_{i0}}{\left(2\pi v^{2}_{Ti}
\right)^{3/2}}
\exp\left(-\frac{v^{2}_{i}}{v^{2}_{Ti}} \right),
\end{eqnarray*}
the ion density perturbation Fourier transformed over 
coordinate $\mathbf{r}_{i}$ is
\begin{eqnarray}
&\displaystyle n^{(i)}_{i}\left(\mathbf{k}_{i},t \right)=-\frac{en_{0i}}{T_{i}}\int
\limits^{t}_{t_{0}}dt_{1}\varphi_{i}\left(\mathbf{k}
_{i},t_{1} \right)k^{2}_{i}v^{2}_{Ti}\left(t-t_{1}\right)e^{-\frac{1}{2}k^{2}_{i}v^{2}_{Ti}
\left(t-t_{1}\right)^{2}}.\label{19}
\end{eqnarray}

The temporal evolution of the separate spatial harmonic of the potential 
$\varphi$  with Poisson equation (\ref{3}) may be investigated  in the electron frame as the 
equation for $\varphi_{e}\left(\mathbf{k}_{e},t\right)$ by the Fourier transform of Eq. 
(\ref{3}) over $\mathbf{r}_{e}$, or as the equation for  $\varphi_{i}\left(\mathbf{k}_{i},t
\right)$ by the Fourier transform of Eq. (\ref{3}) over $\mathbf{r}_{i}$. For the deriving 
the Fourier transformed Poisson equation (\ref{3}) for $\varphi_{e}\left(\mathbf{k}_{e},t
\right)$ the Fourier transform over $\mathbf{r}_{e}$ should be determined for $n_{i}
\left(\mathbf{r}_{i},t\right)$ and for potential $\varphi_{i}\left(\mathbf{r}_{i},t_{1} 
\right)$. With coordinates transform (\ref{8}) we obtain, 
that
\begin{eqnarray}
&\displaystyle \int d\mathbf{r}_{e} n_{i}\left(\mathbf{r}_{i},t \right)e^{-i\mathbf{k}_{e}
\mathbf{r}_{e}}=n^{(e)}_{i}\left(\mathbf{k}
_{e},t \right)
\nonumber
\\ &\displaystyle
= n^{(i)}_{i}\left(k_{ex}-k_{ey}V'_{0}t, k_{ey}, k_{ez}, t \right)e^{ik_{ey}V_{0}t}.
\label{20}
\end{eqnarray}
where the superscript in $n^{(e)}_{i}$ denotes that the Fourier transform of $n_{i}$ is 
calculated in the electron convected frame.

This relation means that for obtaining the Fourier transform for the ion density perturbation 
$n^{(i)}_{i}\left(\mathbf{k}
_{i},t \right)$ over the coordinates $\mathbf{r}_{e}$, it is necessary to multiple $n^{(i)}
_{i}\left(\mathbf{k}_{i},t \right)$ on 
$e^{ik_{ey}V_{0}t}$, that corresponds to the known Doppler effect for the frames which move 
with relative steady uniform velocity $
\mathbf{V}_{0}\parallel \mathbf{e}_{y}$,  and to change  the components of the wavevector $
\mathbf{k}_{i}$ in $n^{(i)}_{i}
\left(\mathbf{k}_{i}, t\right)$ on the components of the wavevector $\mathbf{k}_{e}$ as 
\begin{eqnarray}
&\displaystyle k_{ix}=k_{ex}-k_{ey}V'_{0}t, \quad k_{iy}=k_{ey}, \quad k_{iz}=k_{ez}.
\label{21}
\end{eqnarray}
 
Equation (\ref{19}) for $ n^{(i)}_{i}\left(\mathbf{k}_{i},t \right)$ includes the potential 
$\varphi_{i}\left(\mathbf{k}_{i},t_{1}\right)$. Using the relation
\begin{eqnarray}
&\displaystyle \varphi_{i}\left(\mathbf{k}_{i},t_{1}\right)=\int d\mathbf{r}_{i}\varphi_{i}
\left(\mathbf{r}_{i},t_{1}\right)e^{-i\mathbf{k}_{i}\mathbf{r}_{i}}
\nonumber
\\ &\displaystyle
= \varphi_{e}\left(k_{ix}+k_{iy}V'_{0}t_{1}, k_{iy}, k_{iz},t_{1}\right)e^{-ik_{iy}V_{0}
t_{1}},\label{22}
\end{eqnarray}
which follows from the identity $\varphi_{i}\left(\mathbf{r}_{i},t_{1}\right)=\varphi_{e}
\left(\mathbf{r}_{e},t_{1}\right)$, and relations (\ref{21}), we 
find that potential  $\varphi_{i}\left(\mathbf{k}_{i},t_{1}\right)$ in Eq. (\ref{19}) should 
be changed on the $\varphi_{e}$ by employing the identity
\begin{eqnarray}
&\displaystyle \varphi_{i}\left(\mathbf{k}_{i},t_{1}\right)
\nonumber
\\ &\displaystyle
=\varphi_{e}\left(k_{ex}-k_{ey}V'_{0}\left(t-t_{1}\right), k_{ey}, k_{ez}, t_{1}\right)e^{-
ik_{ey}V_{0}t_{1}}.\label{23}
\end{eqnarray}

Relations (\ref{20}) and (\ref{23}) demonstrate that the separate spatial Fourier mode of the 
ion density perturbation $n^{(i)}_{i}\left(\mathbf{k}_{i},t 
\right)$ and potential $\varphi_{i}\left(\mathbf{r}_{i},t_{1}\right)$ determined in the ion 
frame are detected in the electron frame as \textit{the Doppler-shifted continuously sheared 
modes with time-dependent wave numbers}. Equation (\ref{20}) displays that the time-dependent 
non-modal effect of the flow shear becomes important at time $t$ for which $|k_{ey}V'_{0}t|
\geq |k_{ex}|$. For $k_{ey}\sim k_{ex}$ and time $t\sim \gamma^{-1}$, where $\gamma$ is the 
growth rate of the considered modal instability, the separate spatial mode of the ion density 
perturbation is observed in the electron frame as a  non-modal structure changed with time 
when $V'_{0}\gtrsim \gamma$.

Employing connection relations (\ref{20}) and (\ref{23}) in Eq. (\ref{19}) for $n^{(i)}_{i}
\left(\mathbf{k}_{i},t \right)$, we obtain
the equation governing the temporal evolution of the potential $\varphi_{e}\left(\mathbf{k}
_{e},t\right)$ 
in the Hall plasma with a sheared Hall current,
\begin{widetext}
\begin{eqnarray}
&\displaystyle K^{2}_{e}\left(t\right)\varphi_{e}\left(\mathbf{k}_{e},t\right)
\nonumber
\\ &\displaystyle
=-\frac{1}{\lambda^{2}_{De}}\sum\limits^{\infty}_{n=-\infty}\int\limits^{t}_{t_{0}}dt_{1}
\varphi_{e}\left(\mathbf{k}_{e}, t_{1}\right)
\left(in\omega_{ce}+k_{ez}^{2}v^{2}_{Te}\left(t-t_{1}\right)\right)
\nonumber
\\ &\displaystyle
\times A_{en}\left(t, t_{1}\right)e^{-\frac{1}{2}k^{2}_{ez}v^{2}_{Te}\left(t-
t_{1}\right)^{2}-in\omega_{ce}\left(t-t_{1}\right)-in\left(\theta\left(t\right)-\theta
\left(t_{1}\right)\right)} 
\nonumber
\\ &\displaystyle
-\frac{1}{\lambda^{2}_{Di}}\int\limits^{t}_{t_{0}}dt_{1} \varphi_{e}\left(k_{ex}-k_{ey}V'_{0}
\left(t-t_{1}\right), k_{ey}, k_{ez}, t_{1}\right)e^{ik_{ey}
V_{0}\left(t-t_{1}\right)}
\nonumber
\\ &\displaystyle
\times K^{2}_{e}\left(t\right)v^{2}_{Ti}\left(t-t_{1}\right) 
\exp\left[-\frac{1}{2}K^{2}_{e}\left(t\right)v^{2}_{Ti}\left(t-t_{1}\right)^{2}
\right],
\label{24}
\end{eqnarray}
where 
\begin{eqnarray}
&\displaystyle
A_{en}\left(t, t_{1}\right)= I_{n}\left(k_{e\bot}\left(t\right)k_{e\bot}\left(t_{1}\right)
\rho_{e}^{2}\right)e^{ -\frac{1}{2}\rho_{e}^{2}\left(k_{e\bot}^{2}\left(t\right)+k^{2}_{e
\bot}\left(t_{1}\right)\right)}
\label{25}
\end{eqnarray}
with $K^{2}_{e}\left(t\right)=\left(k_{ex}-k_{ey}V'_{0}t \right)^{2}+ k^{2}_{ey} + k_{ez}^{2}
$;  $\lambda_{Di,e} = \left(T_{i,e}/4\pi n_{i0.e0}e^{2}\right)^{1/2}$  is the 
ion, electron Debye length. The counterpart of this equation - the equation for $\varphi_{i}
\left(\mathbf{k}_{i},t\right)$ has a form
\end{widetext}
\begin{eqnarray}
&\displaystyle k^{2}_{i}\varphi_{i}\left(\mathbf{k}_{i},t\right)=-\frac{1}{\lambda^{2}_{De}}
\sum\limits^{\infty}_{n=-\infty}\int\limits^{t}_{t_{0}}dt_{1}
\varphi_{i}\left(k_{ix}+k_{iy}V'_{0}\left(t-t_{1}\right), k_{iy}, k_{iz}, t_{1}\right)
\nonumber
\\ &\displaystyle
\times e^{-ik_{iy}V_{0}\left(t-t_{1}\right)-in\omega_{ce}\left(t-t_{1}\right)-in\left(\theta
\left(t\right)-\theta\left(t_{1}\right)\right)}\left(in
\omega_{ce}+k^{2}_{iz}v^{2}_{Te}\left(t-t_{1}\right) \right) 
\nonumber
\\ &\displaystyle
\times A_{en}\left(t, t_{1}\right)\exp\left[-\frac{1}{2}k^{2}_{iz}v^{2}_{Te}\left(t-t_{1}
\right)^{2}\right]
\nonumber
\\ &\displaystyle
-\frac{1}{\lambda^{2}_{Di}}\int\limits^{t}_{t_{0}}dt_{1}\varphi_{i}\left(\mathbf{k}_{i}, 
t_{1}\right)k^{2}_{i}v^{2}_{Ti}\left(t-t_{1}\right) 
e^{-\frac{1}{2}k^{2}_{i}v^{2}_{Ti}\left(t-t_{1}\right)^{2}}.
\label{26}
\end{eqnarray}
in which the relations 
\begin{eqnarray}
&\displaystyle 
n_{e}^{(i)}\left(\mathbf{k}_{i}, t\right)=n_{e}^{(e)}\left(k_{ix}+V'_{0}tk_{iy},k_{iy}, 
k_{iz}, t\right)e^{-ik_{iy}V_{0}t}
\label{27}
\end{eqnarray}
and 
\begin{eqnarray}
&\displaystyle \varphi_{e}\left(\mathbf{k}_{e},t_{1}\right)
\nonumber
\\ &\displaystyle
=\varphi_{i}\left(k_{ix}+k_{iy}V'_{0}\left(t-t_{1}\right), k_{iy}, k_{iz}, t_{1}
\right)e^{ik_{iy}V_{0}t_{1}}
\label{28}
\end{eqnarray}
were used.

For the spatially homogeneous  electric field $\mathbf{E}_{0}$ for which $V'_{0}=0$, Eqs. 
(\ref{20}), (\ref{21}), and (\ref{23}) reproduce the known relations for the Doppler effect:
\begin{eqnarray}
&\displaystyle 
n^{(e)}_{i}\left(\mathbf{k}
_{e},t \right)= n^{(i)}_{i}\left(\mathbf{k}_{e}, t \right)e^{ik_{ey}V_{0}t},
\label{29}
\\
&\displaystyle \mathbf{k}_{i}= \mathbf{k}_{e},
\label{30}
\\
&\displaystyle \varphi_{i}\left(\mathbf{k}_{i},t_{1}\right)
=\varphi_{e}\left(\mathbf{k}_{e}, t_{1}\right)e^{-ik_{ey}V_{0}t_{1}}.\label{31}
\end{eqnarray}
In this case of the uniform Hall current, or when the local approximation (\ref{64}) for the 
inhomogeneous current velocity is applied for which the inhomogeneous velocity is considered 
as almost uniform and the velocity shear does not distinguish,  Eqs. (\ref{24}) and 
(\ref{26}) are the integral equations of the convolution
type, which can be solved by using various kinds of integral transforms. In the $t_{0}
\rightarrow -\infty$ limit, 
Eqs. (\ref{24}) and (\ref{26}) have normal modes solutions for the Fourier transformed over 
time potentials in the form 
\begin{eqnarray}
&\displaystyle 
\varphi_{e}\left(\mathbf{k}_{e}, \omega_{e}\right)\left(1+\varepsilon_{e}\left(\mathbf{k}
_{e}, \omega_{e}\right)+ \varepsilon_{i}\left(\mathbf{k}_{e}, 
\omega_{e}+\mathbf{k}_{e}\mathbf{V}_{0}\right)\right)=0
\label{32}
\end{eqnarray}
for Eq.(\ref{24}), and solution 
\begin{eqnarray}
&\displaystyle 
\varphi_{i}\left(\mathbf{k}_{i}, \omega_{i}\right)\left(1+\varepsilon_{i}\left(\mathbf{k}
_{i}, \omega_{i}\right)+ \varepsilon_{e}\left(\mathbf{k}_{i}, 
\omega_{i}-\mathbf{k}_{i}\mathbf{V}_{0}\right)\right)=0
\label{33}
\end{eqnarray}
for Eq. (\ref{26}), where $\varepsilon_{i}$ and $\varepsilon_{e}$ are known ion and electron 
components of the electrostatic dielectric permittivity of the 
Hall plasma. In the cases of the sheared Hall current, the solutions to Eqs. (\ref{24}),  
(\ref{26}) can't be presented in the modal forms (\ref{32}) or (\ref{33}), and the solution 
of Eqs. (\ref{24}), (\ref{26}) as the initial value problems are necessary. 

It is obvious that the exact analytical solutions to integral equations (\ref{24}) and 
(\ref{26}) cannot be obtained explicitly. In this paper, we present the  approximate nonmodal 
solutions of the integral equations (\ref{24}) and (\ref{26}) for two basic classes of 
instabilities: reactive  (MTS instability)  in Sec. \ref{sec3} and kinetic (IS instability), 
in Sec. \ref{sec4}. The solutions are obtained for the case of the velocity shear $V'_{0}$ 
much less than the frequency $\omega_{0}$ of the corresponding modal instability of the 
shearless Hall plasma. The effect of such current shear is relatively small at the time 
interval $t-t_{0}$ for which 
\begin{eqnarray}
&\displaystyle 
\omega_{0}^{-1}\ll t-t_{0}< \left(V'_{0}\right)^{-1},
\label{34}
\end{eqnarray}
The modal theory of the MTS instability of the Hall plasma with  uniform current employs 
the approximation of the hydrodynamic ions and electrons which assumes that the thermal 
velocities of ion and electrons are much less than the phase velocity of the unstable 
perturbations. In Sec. \ref{sec3}, we present simple procedure for the deriving the  
analytical nonmodal solution to Eq. (\ref{24}) for the  
modified two-stream instability in this hydrodynamic approximation with accounting for the 
effect of the weak velocity shear.  

Other model, of the hydrodynamic ions, but of adiabatic electrons the thermal velocity of 
which is larger than the phase velocity of the unstable perturbations, is employed  for the 
modal IS kinetic instability of a plasma with uniform current. In Sec. \ref{sec4}, we 
derive for this model the solution to Eq. (\ref{26}) for the nonmodal kinetic IS 
instability for the case (\ref{34}) of the weak current velocity shear.

\section{The hydrodynamic nonmodal modified two-stream instability}\label{sec3}

In this section, we consider the temporal evolution of the MTS 
instability\cite{Krall, Lashmore-Davies} of the Hall plasma
with a sheared Hall current. This investigation may be performed using any of Eqs. (\ref{24}) 
and (\ref{26}). Here we employ for this task Eq. (\ref{24}). 

The MTS instability is a long-wavelength, $k_{\bot}\rho_{e} \ll 1$, instability 
which develops in the intermediate-frequency range  $\omega_{ci}\ll 
\omega\ll \omega_{ce}$ in plasmas with electrons drifting relative to ions across the 
magnetic field. The phase velocity across the magnetic field of the 
unstable waves of MTS instability is much above the ion thermal velocity, and the phase 
velocity along the magnetic field is much above the electron 
thermal velocity. The dispersion equation for this instability in the electron frame is
\begin{eqnarray}
&\displaystyle
1+\frac{\omega_{pe}^{2}}{\omega^{2}_{ce}}-\frac{k^{2}_{ez}}{k^{2}_{ey}}\frac{\omega^{2}_{pe}}
{\omega^{2}\left(\mathbf{k}_{e}\right)}-\frac{\omega^{2}_{pi}}
{\left(\omega\left(\mathbf{k}_{e}\right)+k_{ey}V_{0}\right)^{2}}=0.\label{35}
\end{eqnarray}
The MTS instability is an example of a general class of instabilities which have been 
referred to as reactive\cite{Hasegawa, Cap}. These instabilities occurs when two wave modes 
couple at a critical frequency. The MTS instability develops due to the coupling
\cite{Lashmore-Davies} the electron mode with the frequency $\omega_{1}=\omega_{Lh}
\left(k_{z}/k\right)\left(m_{i}/m_{e}\right)^{1/2}
$ with Doppler-shifted lower hybrid wave $\omega_{2}=\omega_{Lh}-k_{y}V_{0}$, where $
\omega_{Lh}$ is the lower hybrid frequency,
\begin{eqnarray*}
&\displaystyle
\omega^{2}_{Lh}=\omega^{2}_{pi}\left(1+\frac{\omega^{2}_{pe}}{\omega^{2}_{ce}}\right)^{-1}.
\end{eqnarray*}
under conditions when the frequencies $\omega_{1}$ and $\omega_{2}$ are almost equal.
The solution of Eq. (\ref{35}), 
\begin{eqnarray}
&\displaystyle 
\omega\left(\mathbf{k}_{e}\right)=\frac{1}{2}\left(\omega_{Lh}-k_{ey}V_{0}-\omega_{Lh}
\frac{k_{ez}}{k_{e}}\left(\frac{m_{i}}{m_{e}}\right)^{1/2}\right)
\nonumber
\\ &\displaystyle
\pm \frac{1}{2}\left[\left(\omega_{Lh}-k_{ey}V_{0}+\omega_{Lh}\frac{k_{ez}}{k_{e}}
\left(\frac{m_{i}}{m_{e}}\right)^{1/2}\right)^{2}-\omega^{2}_{Lh}
\frac{k_{ez}}{k_{e}}\left(\frac{m_{i}}{m_{e}}\right)^{1/2} \right]^{1/2} ,\label{36}
\end{eqnarray}
predicts the MTS instability development for $k_{ez}/k_{e} \sim \left(m_{e}/m_{i}\right)^{1/
2}$.

For the solution of the integral equation (\ref{24}) for the instabilities which have the 
phase velocities larger than the thermal velocities of particles, 
the alternative form of Eq. (\ref{24}), resulted from the integration by parts of Eq. 
(\ref{24}), is more suitable,
\begin{eqnarray}
&\displaystyle \left(K_{e}^{2}\left(t\right)+\frac{1}{\lambda^{2}_{De}}+\frac{1}{\lambda^{2}
_{Di}}\right)\varphi_{e}\left(\mathbf{k}_{e},t\right)
\nonumber
\\ &\displaystyle
=\frac{1}{\lambda^{2}_{De}}\sum
\limits^{\infty}_{n=-\infty}\int\limits^{t}_{t_{0}}dt_{1}\frac{d}{dt_{1}}\left\lbrace 
\varphi_{e}\left(\mathbf{k}_{e}, t_{1}\right)A_{en}\left(t, t_{1}\right)\right.
\nonumber
\\ &\displaystyle
\left.\times e^{-in\omega_{ce}\left(t-t_{1}\right)-in
\left(\theta\left(t\right)-\theta\left(t_{1}\right)\right)}\right\rbrace e^{-\frac{1}{2}k^{2}
_{ez}v^{2}_{Te}\left(t-t_{1}\right)^{2}} 
\nonumber
\\ &\displaystyle
-\frac{1}{\lambda^{2}_{De}}\sum\limits^{\infty}_{n=-\infty}n\omega_{ce}\int\limits^{t}
_{t_{0}}dt_{1}
\varphi_{e}\left(\mathbf{k}_{e}, t_{1}\right)A_{en}\left(t, t_{1}\right)
\nonumber
\\ &\displaystyle
\times e^{-\frac{1}{2}k^{2}_{ez}v^{2}_{Te}\left(t-
t_{1}\right)^{2}-in\omega_{ce}\left(t-t_{1}\right)-in\left(\theta\left(t\right)-\theta
\left(t_{1}\right)\right)}
\nonumber
\\ &\displaystyle
+\frac{1}{\lambda^{2}_{Di}}\int\limits^{t}_{t_{0}}dt_{1}\frac{d}{dt_{1}}\left\lbrace 
\varphi_{e}\left(k_{ex}-k_{ey}V'_{0}\left(t-t_{1}\right), k_{ey}, 
k_{ez}, t_{1}\right)e^{ik_{ey}V_{0}\left(t-t_{1}\right)}\right\rbrace 
\nonumber
\\ &\displaystyle
\times \exp\left[-\frac{1}{2}K^{2}_{e}\left(t\right)v^{2}_{Ti}\left(t-t_{1}\right)^{2} 
\right]+Q\left(\mathbf{k}_{e}, t, t_{0}\right),
\label{37}
\end{eqnarray}
where 
\begin{eqnarray}
&\displaystyle Q\left(\mathbf{k}_{e}, t, t_{0}\right)=\frac{1}{\lambda^{2}_{De}}\varphi_{e}
\left(\mathbf{k}_{e}, t_{0}\right)\frac{T_{i}}{T_{e}}\sum
\limits^{\infty}_{n=-
\infty}A_{en}\left(t, t_{0}\right)
\nonumber
\\ &\displaystyle
\times e^{-\frac{1}{2}k^{2}_{ez}v^{2}_{Te}\left(t-
t_{0}\right)^{2}-in\omega_{ce}\left(t-t_{0}\right)-in\left(\theta\left(t\right)-\theta
\left(t_{0}\right)\right)}
\nonumber
\\ &\displaystyle
+\frac{1}{\lambda^{2}_{Di}}\varphi_{e}\left(k_{ex}-k_{ey}V'_{0}\left(t-t_{0}\right), k_{ey}, 
k_{ez}, t_{0}\right)e^{ik_{ey}V_{0}\left(t-t_{0}\right)}
\nonumber
\\ &\displaystyle
\times e^{-\frac{1}{2}v^{2}_{Ti}\left(\left(k_{ex}-k_{ey}V'_{0}t \right)^{2}+ k^{2}_{ey} + 
k_{ez}^{2} \right)v^{2}_{Ti}\left(t-t_{0}\right)^{2}}
\label{38}
\end{eqnarray}
determines the input from the $t=t_{0}$ limit of the integration of Eq. (\ref{24}) by parts.
The approximations 
\begin{eqnarray}
&\displaystyle 
\exp\left(-\frac{1}{2}k^{2}_{ez}v^{2}_{Te}\left(t-t_{1}\right)^{2}\right)\approx 1-\frac{1}
{2}k^{2}_{ez}v^{2}_{Te}\left(t-t_{1}\right)^{2},
\nonumber
\\ &\displaystyle
\exp\left(-\frac{1}{2}K^{2}_{e}\left(t\right)v^{2}_{Ti}\left(t-t_{1}\right)^{2}\right)\approx 
1-\frac{1}{2}K^{2}_{e}\left(t\right)v^{2}_{Ti}\left(t-t_{1}
\right)^{2},
\label{39}
\end{eqnarray} 
which corresponds to the weak electron and ion Landau damping (hydrodynamic approximation) 
strongly simplify the solution of Eq. (\ref{37}).
Accounting for the only term with $n=0$ in Eq. (\ref{37}) and using the approximation $A_{e0}
\left(t, t_{1}\right)\approx 1$ that is sufficient for a long-wavelength, $k_{\bot}\rho_{e} 
\ll 1$, MTS instability,  which has the frequency and the growth rate much less than the 
electron cyclotron frequency $\omega_{ce}$, we obtain the equation
\begin{eqnarray}
&\displaystyle \left(K^{2}_{e}\left(t\right)+ \frac{1}{\lambda^{2}_{De}}+\frac{1}{\lambda^{2}
_{Di}}\right)\varphi_{e}\left(\mathbf{k}_{e}, t\right)
\nonumber
\\ &\displaystyle
+\frac{1}{\lambda^{2}_{De}}\int\limits^{t}_{t_{0}}dt_{1}\left(1-\frac{1}{2}k^{2}_{ez}v^{2}
_{Te}\left(t-t_{1}\right)^{2}\right)\frac{d}{dt_{1}} \varphi_{e}
\left(\mathbf{k}_{e}, t_{1}\right)
\nonumber
\\ &\displaystyle
+\frac{1}{\lambda^{2}_{Di}}\int\limits^{t}_{t_{0}}dt_{1}\frac{d}{dt_{1}}\left\lbrace 
\varphi_{e}\left(k_{ex}-k_{ey}V'_{0}\left(t-t_{1}\right), k_{ey}, 
k_{ez}, t_{1}\right)e^{ik_{ey}V_{0}\left(t-t_{1}\right)}\right\rbrace 
\nonumber
\\ &\displaystyle
\times \left(1-\frac{1}{2}K^{2}_{e}v^{2}_{Ti}\left(t-t_{1}\right)^{2}\right)
=Q\left(\mathbf{k}_{e}, t, t_{0}\right).
\label{40}
\end{eqnarray}
We will find the solution of Eq. (\ref{40}) under condition (\ref{34}) in the WKB-like form
\cite{Mikhailenko1} 
\begin{eqnarray}
&\displaystyle 
\varphi\left(\mathbf{k}_{e}, t_{1} \right) =\Phi_{e}\left(\mathbf{k}_{e} \right)
e^{-i \int\limits^{t_{1}}_{t_{0}}\omega\left(\mathbf{k}_{e}, t_{2} \right)dt_{2}}, 
\label{41}
\end{eqnarray} 
where $\Phi\left(\mathbf{k}_{e}\right)=\int \limits^{\infty}_{-\infty}e^{-i\mathbf{k}_{e}
\mathbf{r}_{e}}\varphi\left(\mathbf{k}_{e}, t_{0}\right)
d\mathbf{k}_{e}$ is the Fourier transform of the initial perturbation of $\varphi
\left(\mathbf{r}_{e}, t_{1} \right)$ at $ t_{1}=t_{0}$. 
The equation for $\omega\left(\mathbf{k}_{e}, t\right)$ is derived iteratively by integration 
by parts of Eq. (\ref{40})  in the form of a power series 
expansion in powers of $|V'_{0}/\omega_{0}|<1$, 
\begin{eqnarray}
&\displaystyle
\Phi_{e}\left(\mathbf{k}_{e} \right)
e^{-i \int\limits^{t}_{t_{0}}\omega\left(\mathbf{k}_{e}, t_{1} \right)dt_{1}}\left[1+
\frac{\omega_{pe}^{2}}{\omega^{2}_{ce}}-\frac{k^{2}_{ez}}{k^{2}
_{ey}}\frac{\omega^{2}_{pe}}{\omega^{2}_{0}\left(\mathbf{k}_{e}\right)}
\left(1+\frac{3i}{\omega_{0}^{2}\left(\mathbf{k}_{e}\right)}\frac{d\omega\left(\mathbf{k}
_{e}, t\right)}{dt}\right)\right.
\nonumber
\\ &\displaystyle
\left.-\frac{\omega^{2}_{pi}}{\left(\omega_{0}\left(\mathbf{k}_{e}\right)+k_{ey}V_{0}
\right)^{2}}\left(1-2\frac{k_{ey}V'_{0}}{\left(\omega_{0}
\left(\mathbf{k}
_{e}\right)+k_{ey}V_{0}\right)}\frac{\partial \omega_{0}\left(\mathbf{k}_{e}\right)}{\partial 
k_{ex}}\left(t-t_{0}\right)\right)\right]= Q\left(\mathbf{k}_{e}, t, t_{0}\right),
\label{42}
\end{eqnarray} 
where $\omega_{0}\left(\mathbf{k}_{e}\right)$ is the solution (\ref{36}) of Eq. (\ref{35}). 
For the potential exponentially growing with time, we can 
neglect by $Q\left(\mathbf{k}_{e}, t, t_{0}\right)$ in Eq. (\ref{40}) and obtain the solution 
for the exponential of Eq. (\ref{41}),
\begin{eqnarray}
&\displaystyle
-i\int\limits^{t}_{t_{0}}\omega\left(\mathbf{k}_{e}, t_{1} \right)dt_{1}=-i\omega_{0}
\left(\mathbf{k}_{e}\right)\left(t-t_{0}\right)
\nonumber
\\ &\displaystyle
+\frac{1}{9}\omega^{2}_{Lh}
\left(\frac{m_{i}}{m_{e}}\right)^{3/2}\left(\frac{k_{ez}}{k_{e}}\right)^{3}\frac{k_{ex}}
{k^{2}_{e}}k_{ey}V'_{0}\left(t-t_{0}\right)^{3}.
\label{43}
\end{eqnarray} 
This solution predicts fast nonmodal growth as $\exp a\left(t-t_{0}\right)^{3}$ of the 
potential $\varphi_{e}\left(\mathbf{k}_{e},t\right)$ for the perturbations with $k_{ey}V'_{0}
>0$. 

Equation (\ref{43}) displays that the uniform and the sheared components of the current 
velocity are the independent sources of the current driven instabilities. We found that the
uniform part, $V_{0}$, of the current velocity is responsible for the modal type of the 
instability development, whereas the current velocity shear $V'_{0}$ is the source of the 
free energy for the development of the instability of the nonmodal type. 
\section{The nonmodal kinetic ion-sound instability}\label{sec4}

For $k_{ez}/k_{e}> \left(m_{e}/m_{i}\right)^{1/2}$  the electron mode does not couple with 
the Doppler-shifted lower hybrid wave. In this case, the lower hybrid 
wave goes into the IS wave\cite{Lashmore-Davies,Gary1}. The dispersive properties 
of the long wavelength, $k_{\bot}\rho_{e}\ll 1$, IS 
instability of a plasma with a uniform Hall current are determined by the equation
\begin{eqnarray}
&\displaystyle
\varepsilon_{0}\left(\mathbf{k}, \omega\right)=1-\frac{\omega^{2}_{pi}}{\omega^{2}}+\frac{1}
{k^{2}\lambda^{2}_{De}}+i\frac{1}{k^{2}\lambda^{2}_{De}}\sqrt{2}z_{e}W
\left(z_{e}\right)=0,
\label{44}
\end{eqnarray}
where $W\left(z_{e}\right)=e^{ - z_{e}^{2}}\left(1 +\left(2i / \sqrt {\pi 
}\right)\int\limits_{0}^{z_{e}} e^{t^{2}}dt \right)$ is the complex error function (also  
known as the Faddeeva function\cite{Faddeyeva})with argument $z_{e}=\left(\omega-k_{y}V_{0}
\right)/\sqrt{2}k_{z}v_{Te}
$. The solution of Eq. (\ref{44}) for the adiabatic electrons $\left(|z_{e}|\ll 1\right)$ is 
$\omega\left(\mathbf{k}\right)=\omega_{IS}+\delta \omega\left(\mathbf{k}\right)$, where $
\omega_{IS}\left(\mathbf{k}\right)$ is the frequency of the ion 
sound wave, $\omega^{2}_{IS}\left(\mathbf{k}\right)=k^{2}v^{2}
_{s}\left(1+k^{2}\lambda^{2}_{De}\right)^{-1}$, $v^{2}_{s}=T_{e}/m_{i}$, and 
\begin{eqnarray}
&\displaystyle \delta\omega\left(\mathbf{k}\right)=-\frac{i}{2}\omega_{IS}z_{e0}W\left(z_{e0}
\right)\ll \omega_{IS}\left(\mathbf{k}\right)
\label{45}
\end{eqnarray}
with $z_{e0}=\left(\omega_{IS}\left(\mathbf{k}\right)-k_{y}V_{0}\right)/\sqrt{2}k_{z}v_{Te}$. 
The IS instability develops when $k_{y}V_{0}>kv_{s}$ with 
the growth rate $\gamma_{IS}\left(\mathbf{k}\right)=\text{Im}\,\delta\omega\left(\mathbf{k}
\right)$ and with $|z_{e0}|< 1$ 
when $k_{z}/k>\left(m_{e}/m_{i}\right)^{1/2}$.

Now we consider the temporal evolution of the IS instability in a plasma with a sheared Hall 
current. For this goal, we obtain the solution to Eq. (\ref{26}) for the  potential $
\varphi_{i}\left(\mathbf{k}_{i}, t\right)$ determined in the ion frame. This solution we 
shall find under condition (\ref{34}) in the WKB form 
\begin{eqnarray}
&\displaystyle 
\varphi\left(\mathbf{k}_{i}, t \right) =\Phi_{i}\left(\mathbf{k}_{i} \right)
e^{-i \int\limits^{t}_{t_{0}}\omega\left(\mathbf{k}_{i}, t_{1} \right)dt_{1}}. 
\label{46}
\end{eqnarray} 
For the deriving the equation for the frequency $\omega\left(\mathbf{k}_{i}, t_{1} \right)$ 
with a weak time-dependence resulted from the current velocity 
shearing, we perform the integration by parts in the ion term using the relation,
\begin{eqnarray}
&\displaystyle 
e^{-i\int\limits^{t}_{t_{0}}\omega\left(\mathbf{k}_{i}, t_{1}\right)dt_{1}}=\frac{i}{\omega
\left(\mathbf{k}_{i}, t \right)} \frac{d}{dt}
\left(e^{-i \int\limits^{t}_{t_{0}}\omega\left(\mathbf{k}_{i}, t_{1} \right)dt_{1}}\right),
\label{47}
\end{eqnarray} 
and derive the expansion of the ion term in the form of the power series of $k_{i}v_{Ti}/
\omega\left(\mathbf{k}_{i}, t\right)$. In the electron term of Eq. 
(\ref{26}), we employ the expansion
\begin{eqnarray}
&\displaystyle
\varphi_{i}\left(k_{ix}+k_{iy}V'_{0}\left(t-t_{1}\right), k_{iy}, k_{iz}, t_{1}\right)
\nonumber
\\ &\displaystyle
= \varphi_{i}\left(\mathbf{k}_{i}, t_{1}\right)+k_{iy}V'_{0}\left(t-
t_{1}\right)\frac{\partial\varphi_{i}}{\partial k_{ix}},
\label{48}
\end{eqnarray}
which is valid under condition (\ref{34}). For the adiabatic electrons, for which $
\omega\ll k_{z}v_{Te}$, the main input into an integral over time $t_{1}$ in 
the electron term of Eq. (\ref{26}) gives the time interval $|t-t_{1}|\lesssim \left(k_{z}
v_{Te}\right)^{-1}$, that defines the validity of  the approximation 
\begin{eqnarray}
&\displaystyle
e^{-i \int\limits^{t}_{t_{1}}\omega\left(\mathbf{k}_{i}, t_{2}\right)dt_{2}}\approx e^{i
\omega\left(\mathbf{k}_{i}, t\right)\left(t-t_{1}\right)}.
\label{49}
\end{eqnarray}
Then, for the exponential term, $\exp\left(-i\int\limits^{t}_{t_{0}}\omega\left(\mathbf{k}
_{i}, t_{1}\right)dt_{1}\right)$, growing with time, we obtain the equation
\begin{eqnarray}
&\displaystyle
e^{-i \int\limits^{t}_{t_{0}}\omega\left(\mathbf{k}_{i}, t_{1}\right)dt_{1}}\left[
\omega_{IS}\left(\mathbf{k}_{i}\right)+\delta \omega\left(\mathbf{k}_{i}\right)-3i
\frac{\omega^{2}_{pi}}
{\omega^{4}_{IS}\left(\mathbf{k}_{i}\right)}\frac{d\omega\left(\mathbf{k}_{i}, t\right)}{dt}
\right.
\nonumber
\\ &\displaystyle \left.+\frac{1}{k^{2}_{i}\lambda^{2}_{De}}\frac{\left(\omega_{IS}
\left(\mathbf{k}_{i}\right)-k_{iy}V_{0}\right)}{k_{iz}v_{Te}}
\frac{k_{iy}V'_{0}}{k_{iz}v_{Te}}\left(i\frac{\partial \ln \Phi_{i}\left(\mathbf{k}_{i} 
\right)}{\partial k_{ix}}+\frac{\partial \omega_{IS}
\left(\mathbf{k}_{i}\right)}{\partial k_{ix}}\left(t-t_{0}\right)\right) \right] =0.
\label{50}
\end{eqnarray}
The solution of this equation is straightforward and gives the following exponential for 
solution (\ref{46}): 
\begin{eqnarray}
&\displaystyle
-i\int\limits^{t}_{t_{0}}\omega\left(\mathbf{k}_{i}, t_{1} \right)dt_{1}=-i\left(\omega_{IS}
\left(\mathbf{k}_{i}\right)+\delta \omega\left(\mathbf{k}_{i}\right)\right)\left(t-t_{0}
\right)
\nonumber
\\ &\displaystyle
-\frac{1}{6}\frac{1}{k^{2}_{i}\lambda^{2}_{De}}\frac{\omega^{4}_{0}\left(\mathbf{k}_{i}
\right)}{\omega^{2}_{pi}}\left(1-k_{i\bot}^{2}\rho^{2}_{e}\right)
\frac{\left(\omega_{IS}\left(\mathbf{k}_{i}\right)-k_{iy}V_{0}\right)}{k_{iz}v_{Te}}
\frac{k_{iy}V'_{0}}{k_{iz}v_{Te}}
\nonumber
\\ &\displaystyle
\times\left(i\frac{\partial \ln \Phi\left(\mathbf{k}_{i}\right)}{\partial k_{ix}}\left(t-
t_{0}\right)^{2}+\frac{1}{3}
\frac{\partial \omega_{IS}\left(\mathbf{k}_{i}\right)}{\partial k_{ix}}\left(t-t_{0}
\right)^{3}\right).
\label{51}
\end{eqnarray}
The first term in the right part of Eq. (\ref{51}) corresponds to the modal IS instability 
evolution with growth rate  $\gamma_{IS}\left(\mathbf{k}\right)$. The second term 
proportional to $V'_{0}$ describes the nonmodal instability. The nonmodal growth is 
determined by the relation
\begin{eqnarray}
&\displaystyle
\int\limits^{t_{1}}_{t_{0}}\gamma_{nm}\left(\mathbf{k}_{i}, t_{2} \right)dt_{2}=\frac{1}{18}
k^{2}_{i}v^{2}_{s}\left(\frac{m_{e}}{m_{i}}\right)^{1/2}\left(1-k_{i\bot}^{2}\rho^{2}_{e}
\right)\frac{k_{iy}k_{ix}}{k_{iz}k_{i}}
\nonumber
\\ &\displaystyle
\times \frac{\left(k_{iy}V_{0}-\omega_{IS}\left(\mathbf{k}_{i}\right)\right)}{k_{iz}v_{Te}}
V'_{0}\left(t-t_{0}\right)^{3}.
\label{52}
\end{eqnarray}
Equation (\ref{52}) displays that the nonmodal growth of the IS perturbations due to the 
shearing of the Hall current occurs for the IS perturbations with   
$k_{iy}V_{0}>\omega_{IS}\left(\mathbf{k}_{i}\right)$ and $\left(k_{ix}/k_{iz}\right)k_{iy}
V'_{0}>0$. The nonmodal growth is independent process from the development of the modal 
instability and accompanies it. In the general case, the instability driven by the current 
with current velocity shear includes modal and nonmodal growth and the net effect of the 
instability development is determined as a balance between them. Eq. (\ref{51}) predicts that 
the nonmodal growth dominates over the modal growth when 
\begin{eqnarray}
&\displaystyle\omega_{IS}\left(\mathbf{k}_{i}
\right)V'_{0}\left(t-t_{0}\right)^{2}>\frac{k_{iz}}{k_{i}}\left(\frac{m_{i}}{m_{e}}
\right)^{1/2}\frac{k^{2}_{i}}{k_{iy}k_{ix}}.
\label{53}
\end{eqnarray}
The nonmodal growth occurs also for the subcritical perturbations for which $\left(k_{ix}/
k_{iz}\right)k_{iy}V'_{0}<0$ with  $k_{iy}V_{0}<\omega_{IS}
\left(\mathbf{k}_{i}\right)$ including the case when $V_{0}=0$. These perturbations are 
suppressed with damping rate determined by Eq. (\ref{45}), but become growing with time when 
condition (\ref{53}) holds. The evolution of the kinetic instability at longer time at which 
condition (\ref{34}) does not hold continues to be nonmodal. However, this evolution can be 
investigated only by the numerical solution of Eqs. (\ref{26})  or (\ref{24}) as the initial 
value problems.

\section{CONCLUSIONS}\label{sec5}
In this paper, the basic equations (Eqs. (\ref{24}) and (\ref{26})) of the kinetic theory of 
the electrostatic instabilities driven by the Hall current with a sheared current velocity 
were derived employing the shearing modes approach. These equations were obtained for the 
case of a Hall current with uniform current velocity shear, $V'_{0}=const$, without 
application of the local approximation and without imposing on the perturbations the 
requirement to have a static structure of the plane wave $\sim \exp\left(i\mathbf{kr}-i\omega 
t\right)$ with prescribed exponential time dependence of the canonical modal form.
The developed theory predicts that the separate spatial Fourier mode of the perturbations in 
the Hall plasma with the inhomogeneous electric field is determined in the frame convected 
with one of the plasma components. The relations (\ref{20}) - (\ref{23}), which are the 
generalization on the sheared current velocity the relations (\ref{29}) - (\ref{31}) of the 
Doppler effect for the uniform current velocity, display that due to the 
different shearing of the ion and electrons flows in the Hall plasma, this mode is detected 
by the second component as the Doppler - shifted continuously sheared mode with time -
dependent wave numbers. This effect of the mode shearing grows continuously with time and the 
interaction of the  plasma components forms the nonmodal time-dependent process which should 
be investigated as the initial value problem.

The nonmodal solutions of the integral equations (\ref{24}) and (\ref{26}) for two basic 
classes of instabilities: reactive (MTS instability) and kinetic (IS instability), are 
obtained in Secs. \ref{sec3} and  \ref{sec4} for the case of the weak uniform current 
velocity shear (\ref{34}) as the solutions of the linear initial value problems. These 
solutions  reveal that the uniform part of the current velocity, $V_{0}$, is responsible for 
the modal evolution of the instability, whereas the current velocity shear, $V'_{0}$, is the 
source of the development of the nonmodal instability with exponent growing with time as $
\sim\left(t-t_{0}\right)^{3}$. This time-dependence, which  is the same as in the solution 
\cite{Mikhailenko1} of the linear initial value problem for the Simon-Hoh 
instability, seems to be common for the plasma instabilities driven by the sheared Hall 
current with uniform shear.

\begin{acknowledgments}
This work was supported by National R$\&$D Program through the National Research Foundation 
of Korea (NRF) funded by 
the Ministry of Education, Science and Technology (Grant No. NRF-2017R1A2B2011106).
\end{acknowledgments}

{}
\end{document}